# Epitaxial growth and the magnetic properties of orthorhombic YTiO$_3$ thin films

S. C. Chae, Y. J. Chang, S. S. A. Seo and T. W. Noh

*ReCOE & FPRD, Department of Physics and Astronomy, Seoul National University, Seoul 151-747, Korea*

D.-W. Kim

*Department of Applied Physics, Hanyang University, Ansan, Gyunggi-do 426-791, Korea*

C. U. Jung[a]

*Department of Physics, Hankuk University of Foreign Studies, Yongin, Gyunggi-do 449-791, Korea*

High-quality YTiO$_3$ thin films were grown on LaAlO$_3$ (110) substrates at low oxygen pressures ($\leq 10^{-8}$ Torr) using pulsed laser deposition. The in-plane asymmetric atomic arrangements at the substrate surface allowed us to grow epitaxial YTiO$_3$ thin films, which have an orthorhombic crystal structure with quite different *a*- and *b*-axes lattice constants. The YTiO$_3$ film exhibited a clear ferromagnetic transition at 30 K with a saturation magnetization of about 0.7 $\mu_B$/Ti. The magnetic easy axis was found to be along the [1–10] direction of the substrate, which differs from the single crystal easy axis direction, i.e., [001].

Recent advances in oxide thin film growth have made it possible to control the oxide film thickness down to the one unit-cell level and to fabricate atomically sharp interfaces.[1] Using epitaxial oxide thin films and heterostructures, numerous workers have discovered intriguing physical phenomena, including strain-induced ferroelectricity, interface magnetism, and structural distortion.[2-6] Elaborate thin film research has contributed greatly to elucidating the underlying mechanisms governing the physical properties at the oxide interfaces; lattice mismatch, stress relaxation, interface chemistry, and polarity discontinuity all play important roles.[7-10]

Quite recently, the interfaces between $LaTiO_3$ and other perovskite titanates have gained growing interest. Ohtomo *et al*. showed that $LaTiO_3/SrTiO_3$ superlattices can have metallic conductivity, although they are composed of two insulating constituents.[2] Okamoto and Millis proposed an electronic reconstruction at the $LaTiO_3/SrTiO_3$ interface: namely, the *d* electrons from the $LaTiO_3$ layers spread across the interface, resulting in the metallic behavior.[11] Subsequently, Ohtomo and Hwang reported that the $LaAlO_3/SrTiO_3$ interface exhibits two-dimensional electron gas behavior.[10]

$YTiO_3$ (YTO) heterostructures are another interesting system. First, YTO is a rare example of a Mott insulator with a ferromagnetic ground state.[12-14] Therefore, if fabricated, the interfaces between YTO and other perovskite titanates could elucidate the role played by magnetism in the electronic reconstruction. Second, since YTO is a ferromagnetic insulating material, it could be used to fabricate multifunctional materials. Recent studies proposed that distortion of Ti-O-Ti bond angle due to a small size of Y ion could allow orbital ordering and stabilize ferromagnetism of YTO.[12] Since such an orbital driven magnetic phase should be highly sensitive to strain states, the growth and characterization of YTO thin films can provide valuable insights into orbital physics.

Despite this importance, few studies have been conducted on YTO thin films, perhaps because growing epitaxial YTO thin films appears challenging. First, the pyrochlore $Y_2Ti_2O_7$ phase is synthesized more easily than the perovskite YTO phase. In bulk synthesis, the YTO phase can be prepared only under a strong reducing atmosphere, e.g., in a mixture of hydrogen and argon.[12] This indicates that a very low oxygen pressure is required to grow single-phase YTO films. Second, YTO has a highly distorted orthorhombic structure with lattice constants of $a$ = 5.679 Å, $b$ = 5.316 Å, and $c$ = 7.611 Å.[14] Many researchers have used cubic substrates with surfaces having a square net, such as SrTiO$_3$ (100) and LaAlO$_3$ (100). On such substrates, the large difference in lattice constants along the $a$- and $b$-axes makes it difficult to grow epitaxial YTO thin films. It would be better to apply anisotropic in-plane stress, which can reduce the large lattice strain at the interface.[8, 15, 16] In this letter, we present structural and magnetic properties of the epitaxial YTO thin films on LaAlO$_3$ (110) substrates. [Lattice constants of LaAlO$_3$ are 5.538 (3.789×√2) Å and 7.578 (3.789×2) Å along the [001] and [1-10] directions, respectively.] YTO thin films were deposited on single-crystal LaAlO$_3$ (LAO) substrates using pulsed laser deposition. The growth mode was monitored carefully using *in situ* reflection high-energy electron diffraction (RHEED).[3, 17] A ceramic Y$_2$Ti$_2$O$_7$ target was ablated using a KrF excimer laser with a fluence of 0.5 J/cm$^2$ at 5 Hz. The partial pressure of oxygen ($P_{O2}$) was varied from 1 × 10$^{-8}$ to 1 × 10$^{-6}$ Torr. The films were grown at substrate temperatures ($T_{sub}$) ranging from 700 to 950ºC, and were cooled slowly to room temperature under a constant $P_{O2}$. The temperature dependence of resistivity (ρ) of YTiO$_3$ thin film with ρ(300K) ~ 30 Ω·cm was similar to that of bulk YTO.[18]

High-quality YTO films grew on the LAO (110) substrates at $P_{O2}$ = 1 × 10$^{-8}$ Torr and $T_{sub}$ = 900ºC. Figure 1(a) shows the x-ray diffraction (XRD) $\theta$–2$\theta$ scan pattern of the YTO film, measured using a synchrotron high-resolution diffractometer at the Pohang Light Source, Korea. The XRD pattern shows only the YTO (200)$_o$ diffraction peak, indicating that the occurrence of negligibly small impurity phases. [In this paper, we choose the pseudo-cubic unit cell notation for LAO and the orthorhombic unit cell notation for YTO. For clarity, we use the subscript 'o' to indicate the orthorhombic unit cell notation.] As shown in the inset, small fringes are observed around the Bragg peak, which likely originated from the interference of the x-rays reflected from both sides of the YTO film. The interference fringes indicate that the film has a uniform thickness and a very smooth surface. From the period of the fringes, the film thickness was estimated to be about 57 nm.[19, 20]

A low oxygen pressure (at least as low as $1 \times 10^{-8}$ Torr) was found to be essential to obtain the desired YTO phase. Figure 1(b) shows the XRD patterns of YTO films fabricated at various $P_{O2}$s. As the $P_{O2}$ increases, the YTO $(200)_o$ peak decreases, while a peak thought to be from the pyrochlore $Y_2Ti_2O_7$ (440) begins to appear. Small peaks from other secondary phases, such as $Y_2TiO_5$, $Y_2Ti_2O_7$, $TiO_2$, and $Ti_2O_3$, were also observed.

Figure 2 shows the RHEED specular beam intensity oscillation obtained during the growth of the YTO film at $P_{O2} = 1 \times 10^{-8}$ Torr. The clear oscillatory behavior indicates that the film was grown in layer-by-layer mode. The inset shows the RHEED pattern after depositing a 50-nm-thick YTO film; the streaky pattern ensures that the grown film has a very flat surface. It was reported that in epitaxial $LaTiO_3$ films, the $La_2Ti_2O_7$ phase could be formed on exposure to air.[21] To test for the possible formation of the $Y_2Ti_2O_7$ phase, we also monitored the evolution of the RHEED patterns after exposing the films to various oxygen pressures from $10^{-8}$ to $10^{-1}$ Torr for 10 min. On increasing the oxygen pressures, the streaks in the patterns gradually weakened and broadened. However, the synchrotron XRD patterns still did not reveal any secondary phase peaks. Although possible formation of a surface layer with other secondary phases in air cannot be ruled out, the amount of such phases should be quite small.

The YTO film was found to be epitaxial and almost fully strained with the substrate. Figures 3(a) and 3(b) show x-ray reciprocal space mappings around the LAO (222) and LAO (400) peaks, respectively. The schematic diagram in Fig. 3(c) shows the in-plane epitaxial relationship between the YTO film and LAO substrate: $[001]_o$(YTO)//[001](LAO) and $[0–10]_o$(YTO)//[1–10](LAO). Note that the horizontal positions of the YTO peak, along the vertical lines in Figs. 3(a) and 3(b), are coincident with those of the LAO substrate, indicating that it should be in the almost fully strained state. From the bulk YTO lattice constant values, it is estimated that the film is under 0.8% (0.4 %) of a tensile (compressive) strain along the $[010]_o$ ($[001]_o$) direction.

On (100)-oriented LAO substrates, we could not get YTO films, which could be well understood by the lattice mismatch. To explain this difference, lattice mismatches and the resulting stress should be considered. The (110)-oriented substrates can induce anisotropic stress. When a YTO film is grown using the configuration in Fig. 3(c), the expected lattice mismatches should be 0.4% along the [001] direction and -0.8% along the [1–10] direction. These values are much smaller than any possible mismatches in the YTO film on (100)-oriented LAO substrates; for example, when the YTO film is grown on LAO (100) substrate aligned in the $[110]_o$ and $[001]_o$ directions along the cubic square net surface, the lattice mismatches should be about 2.7% along [010] and 0.4% along [001]. Therefore, the use of the (110) crystal orientation is beneficial for reducing the strain of highly distorted orthorhombic YTO films on cubic LAO substrates.

To evaluate the magnetic properties of the YTO films, the temperature dependence of magnetization was measured using a SQUID magnetometer. As shown in Fig. 4(a), our YTO films show a clear ferromagnetic transition around 30 K, which is nearly the same as the single crystal value.[13]

Our YTO films exhibited magnetic hysteresis (*M–H*) loops, as shown in Fig. 4(b). The magnetic field was applied along the [1–10] and [001] directions of the LAO (110) substrate. The calculated saturated magnetic moment ($M_s$) of the YTO thin film at 5 K was 0.7 ± 0.05 $\mu_B$/Ti, which is somewhat smaller than that of a single crystal (*i.e.*, 0.83 $\mu_B$/Ti).[13] Tsubota *et al*. reported that the $M_s$ of Y$_{1-x}$Ca$_x$TiO$_3$ single crystals ($x \leq 0.1$) decreased significantly with increasing Ca concentration ($x$), e.g., $M_s$ = 0.5 $\mu_B$/Ti at $x$ = 0.1.[13] In most oxide thin films, lattice mismatch and oxygen nonstoichiometry are unavoidable. The resulting lattice strain and alteration of the Ti valence could reduce the $M_s$ of our film, as the Ca ions do in Y$_{1-x}$Ca$_x$TiO$_3$ single crystals. However, the concentration of oxygen defects should be small, since the $M_s$ of our film (0.7 $\mu_B$/Ti) was similar to that of bulk Y$_{1-x}$Ca$_x$TiO$_3$ with $x$ = 0.02.

In Fig. 4(b), note that the magnetic easy axis of the YTO film is along [0–10]$_o$. When a magnetic field was applied along [100]$_o$, i.e., surface normal the *M–H* loop showed neither saturation, even at a field of 5 T, nor appreciable hysteretic behavior. This was attributable to the shape anisotropy of our thin film geometry. The magnetic anisotropy of YTO single crystals is known to be differ: the easy axis was along the [001]$_o$ direction.[13] In many magnetic oxide thin films including manganites and ruthenates, the substrate-induced strain often changes magnetic easy axes of the thin films.[8, 22, 23] The tensile stress along [010]$_o$ and the compressive stress along [001]$_o$ could possibly make it easier for the spin to be aligned along the [010]$_o$ direction.

In summary, we successfully grew high-quality epitaxial YTiO$_3$ (100) thin films on LaAlO$_3$ (110) substrates. A low $P_{O2}$ ($\leq$10$^{-8}$ Torr) during deposition was crucial for obtaining the desired YTiO$_3$ phase, and the magnetic properties of the thin films except the direction of magnetic easy axes were comparable to the single crystal values.

This work was supported financially by the Korea Science and Engineering Foundation (KOSEF) through the Creative Research Initiative Program and the Center for Strongly Correlated Materials Research and by the Korean Ministry of Education through the BK21 projects. C. U. Jung was also supported by a Korea Research Foundation Grant (MOEHRD; KRF-2005-041-C00165) and by the Basic Research Program of KOSEF (grant no.R01-2006-000-11071-0). Experiments at PLS were supported in part by MOST and POSTECH

a) Corresponding author: cu-jung@hufs.ac.kr

Figure Legends

FIG. 1. (a) XRD $\theta$–$2\theta$ scan patterns of a 57-nm-thick YTiO$_3$ film on LaAlO$_3$ (110) substrate, grown at an oxygen pressure of $10^{-8}$ Torr and substrate temperature of 900ºC. The inset shows a high-resolution scan, indicating interference fringes around the YTO (200)$_o$ peak. (b) XRD $\theta$–$2\theta$ scan patterns of samples deposited under various oxygen partial pressures.

FIG. 2. RHEED intensity oscillation obtained during growth of a YTiO$_3$ film. The inset shows the RHEED pattern after depositing a 50-nm-thick YTO film.

FIG. 3. X-ray reciprocal space mapping around the (a) LaAlO$_3$ (222) and (b) LaAlO$_3$ (400) peaks. (c) A schematic diagram showing the epitaxial growth relationship between the YTO film and the LAO (110) substrate.

FIG. 4. (a) Temperature dependence of magnetization for a YTiO$_3$ film with an applied magnetic field along the [1–10] direction. (b) Magnetic hysteresis loops of the film at 5 K with magnetic field directions along two in-plane directions.

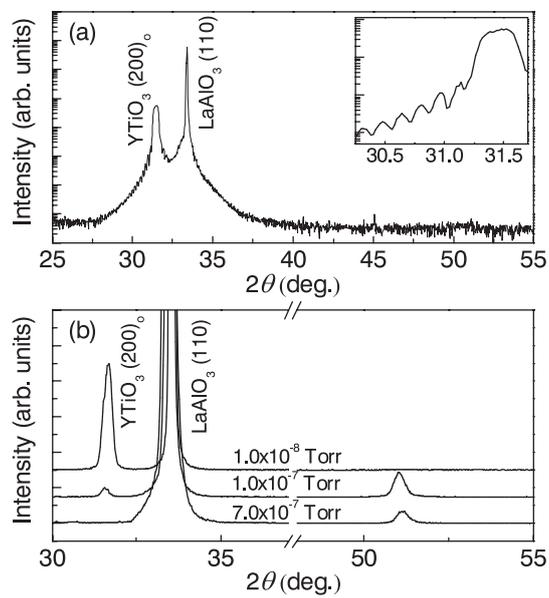

*Figure 1*

S. C. Chae *et al.*

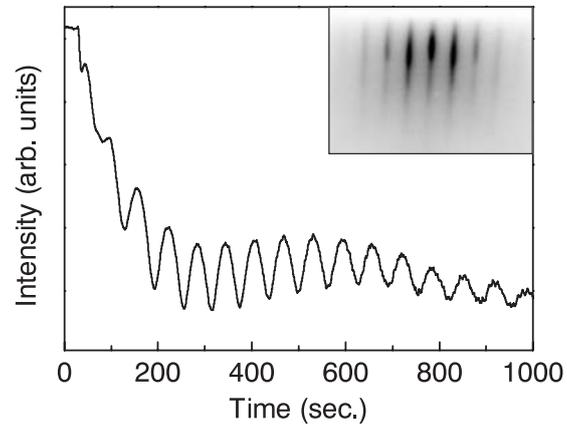

*Figure 2*

S. C. Chae *et al.*

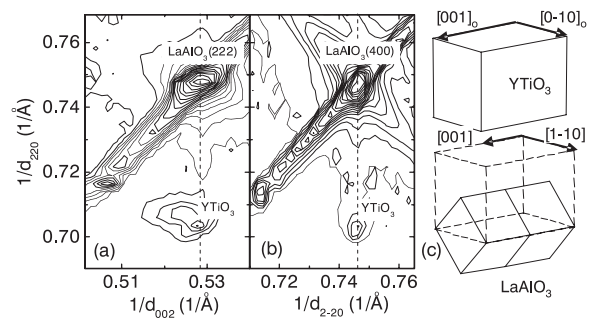

*Figure 3*

S. C. Chae *et al.*

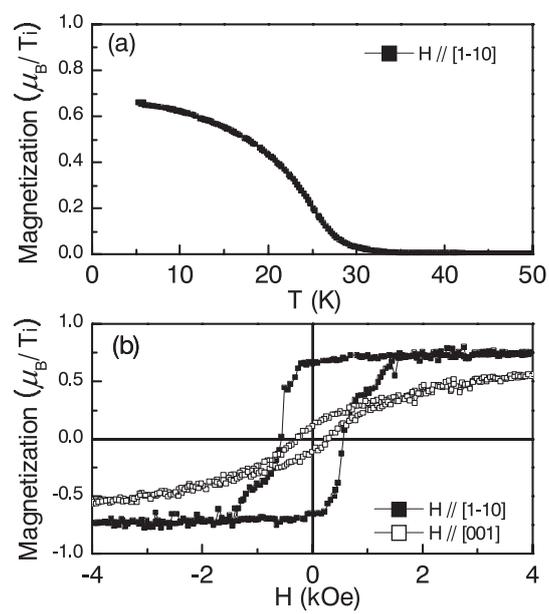

*Figure 4*

S. C. Chae *et al.*